\documentclass{article}
\usepackage{amssymb,amsmath,amsthm}
\usepackage{subcaption,graphicx}
\usepackage{hyperref}
\usepackage{cancel,listings}
\usepackage{comment}
\usepackage[title]{appendix}
\usepackage[superscript]{cite}

\usepackage[margin=2cm]{geometry}
\usepackage[utf8]{inputenc}

\title{Fast post-hoc method for updating moments of large datasets}
\author{B. J. Q. Woods\thanks{E-mail: \href{mailto:benjamin.woods@york.ac.uk}{benjamin.woods@york.ac.uk}}\\ \small{Department of Physics, York Plasma Institute, University of York}, \\ \small{Heslington, York, YO10 5DD, United Kingdom}}
\date{\today}

\begin{document}

\maketitle

\abstract{Moments of large datasets utilise the mean of the dataset; consequently, updating the dataset traditionally requires one to update the mean, which then requires one to recalculate the moment. This means that metrics such as the standard deviation, $R^2$ correlation, and other statistics have to be `refreshed' for dataset updates, requiring large data storage and taking long times to process. Here, a method is shown for updating moments that only requires the previous moments (which are computationally cheaper to store), and the new data to be appended. This leads to a dramatic decrease in data storage requirements, and significant computational speed-up for large datasets or low-order moments (n $\lesssim$ 10).}

\section{Introduction}
This year, the Oak Ridge National Laboratory machine SUMMIT\cite{hines2018stepping} exceeded 1.0 exaops ($10^{18}$ operations per second) computational speed, bringing computing into the exascale era. In this era, scientific simulations will produce larger datasets than before, requiring large amounts of memory and data storage. In the plasma physics community, there has been much recent chatter about the strength of exascale computing and the associated data complications \cite{vay2018warp,smith2018highlights}. Furthermore, the recent increase in popularity of machine learning and deep learning across a variety of different scientific sectors \cite{spears2018deep,eatough2010selection,coudray2018classification} has lead to a growing requirement to produce reduced metrics from datasets in an increasingly efficient fashion.

Computing moments with an updated dataset proves problematic for two reasons: the computational time taken to compute moments can scale with the size of the total data, and one has to retain a large amount of data in storage to compute new moments.

Here, we give an algorithm for computing moments of datasets which is instead computed in time proportional to the size of the data to be appended, and requires only a finite and typically small number of moments to be placed in data storage.

\subsection{Definitions}
We start by stating that we have a dataset $\mathbf{X}$ containing data objects $\{x_i\}$. Let us define the $n^{\textrm{th}}$ moment $M_n[\mathbf{X}]$ as the following:

\begin{equation}
M_n[\{x_i\}] \equiv \dfrac{1}{Z} \sum\limits_{i=1}^{N} f(x_i) (x_i - \bar{x})^n
\label{eq:moment}
\end{equation}

where $\mathbf{X} = \{x_i\}$, and $Z$ is the normalising function, defined here as the sum over $i$ of the weight function $f(x_i)$:

\begin{equation}
Z \equiv \sum\limits_{i=1}^{N} f(x_i)
\end{equation}

Accordingly, $M_0$ is equal to unity. $\bar{x}$ is the weighted mean of the dataset $\mathbf{X}$, such that $M_1$ is identically zero:

\begin{equation}
\bar{x} \equiv \dfrac{1}{Z} \sum\limits_{i=1}^{N} f(x_i) x_i
\end{equation}

\begin{figure}[t!]
    \centering
    \begin{subfigure}[t]{0.45\textwidth}
    \centering
    \includegraphics[width=\textwidth]{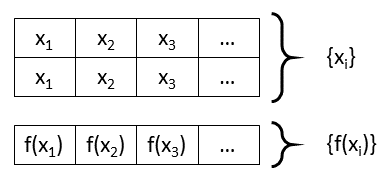}
    \caption{Sketch illustrating the memory occupied by a dataset and a weight function. The data objects here are illustrated to take up twice as much memory for each $i$ as the weight function; as an example, the dataset $\{x_i\}$ could be a set of complex numbers which map onto real scalars via $f$.}
    \label{fig:data_struct}
    \end{subfigure} \hspace{10pt}
    \begin{subfigure}[t]{0.45\textwidth}
    \centering
    \includegraphics[width=0.7\textwidth]{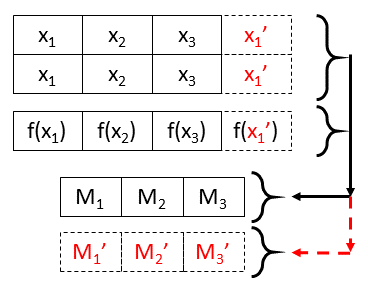}
    \caption{Sketch illustrating an algorithm one might employ to calculate moments of an updated dataset. The new data is append to the dataset, and the moments are calculated again. This requires storing all of the previous data, and is computationally constrained by the size of the full, appended dataset.}
    \label{fig:original_alg}
    \end{subfigure}
    \caption{Sketches showing the memory requirements and algorithmic flow of data in a workflow that produces moments or metrics from a dataset.}
\end{figure}

\subsection{Difficulties with calculating moments}
Firstly, if one computes the original moment (\ref{eq:moment}), this is completed in $\mathcal{O}(N)$ time. This is problematic, because if the dataset is updated $u$ times, then $\mathcal{O}(N)$ time is required on each update:

\begin{equation}
\dfrac{T}{u} = \mathcal{O}(N)
\label{eq:orig_speed}
\end{equation}

where $T$ is the total time taken. The time taken scales linearly with the number of updates, but also with the full dataset size. With data warehouses, this can lead to significant latency issues; in the limit that the increase in dataset size tends to zero, $T/u$ is still finite and non-zero. 

This means that each update requires a full `refresh' which discourages small updates, but also each update requires \emph{more} computational time than any update before it.

Secondly, if one has a dataset which has moments calculated beforehand, the original information is lost; the mapping $\mathbf{X} \to M_n$ is many-to-one, and therefore the mapping is not invertible. This means that one requires an amount of available data storage $S_{\mathbf{X}}$ given by:

\begin{equation}
S_{\mathbf{X}} = \sum\limits_{i=1}^{N} [\textrm{size}(x_i) + \textrm{size}(f(x_i))]
\end{equation}

where $\textrm{size}(x_i)$ is the size of the data $x_i$ and $\textrm{size}(f_i)$ is the size of the data $f(x_i)$. $S_{\mathbf{X}}$ increases with each update, and means that we are required to use larger and large amounts of data to calculate these moments.

However, in principle, the moments themselves take up a much smaller amount of data storage $S_n$:

\begin{equation}
S_n = \textrm{size}(M_n) 
\end{equation}

where $\textrm{size}(M_n)$ is the size of the data required to store each moment. For comparison, if $x_i$, $f(x_i)$ and $M_n$ are datatypes of the same size, then $S_{\mathbf{X}} = 2N S_n$.

\subsection{Dataset operations}
\label{sec:operations}
We begin by defining operations on datasets. To do so, we will take the concept of datasets as a `vector' and generalise the concept of vector space over a field (the field being the data objects, and the vector being the dataset); this will allow us to perform algebra using entire data structures, with a great deal of freedom in the available operations between data.

First, we define a dataset $\mathbf{A}$ as a set of data objects $\{a_i\}$. The dataset is categorized by the type of data objects that form the set, and the size of the dataset. The size of the dataset (cardinality) is referred to as $N$.

It becomes useful to consider the data objects themselves as belonging to a field under addition and multiplication, which we will call the `datatype field'. That is to say, for data objects $p$ and $q$ belonging to a set $\mathbb{G}$ corresponding to a given `datatype':

\begin{equation}
\begin{array}{r c l}
p \circ q \in \mathbb{G} &;& p \times q \in \mathbb{G}
\end{array}
\end{equation}

\begin{figure}[t!]
    \centering
    \includegraphics[width=0.4\textwidth]{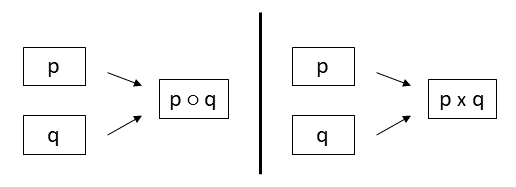}
    \caption{Illustration of the addition and multiplication operators $\circ: \mathbb{G} \times \mathbb{G} \to \mathbb{G}$ and $\times: \mathbb{G} \times \mathbb{G} \to \mathbb{G}$ respectively.}
    \label{fig:operations}
\end{figure}

The addition and multiplication operators ($\circ$ and $\times$) can be defined differently for different datatypes. For example, if we desire that the datatype is a set of cyclic numbers no bigger than 10, one might find:

\[
p \circ q \equiv (p + q) \, \textrm{mod} \, 10 
\]

whereas if the datatype were a set of a complex numbers:

\[
p \circ q \equiv [\Re(p) + \Re(q)] + i[\Im(p) + \Im(q)]
\]

Next, we define the data addition operator between two datasets $\mathbf{A} = \{a_i\}$ and $\mathbf{B} = \{b_i\}$ by the following:

\begin{equation}
\mathbf{A} \circ \mathbf{B} \equiv \{a_i \circ b_i\}
\end{equation}

Formally, the set of datasets $\mathbb{D} \equiv \{\mathbf{A}, \mathbf{B}, \dots\}$ and the data addition operator `$+$' form an Abelian group with identity element $\{0\}$. There is no formal requirement for each member of $\mathbf{A}$ to have the same datatype.

The data multiplication operator between two datasets is defined by the following:

\begin{equation}
\mathbf{A} \times \mathbf{B} \equiv \{a_i \times b_i\}
\end{equation}

The set of datasets $\mathbb{D}$ and the data multiplication operator `$\times$' form an Abelian group with identity element $\{1\}$. As both groups are Abelian, one can now define a mathematical field formed by $\mathbb{D}$, data addition, and data multiplication. Exponentiation is simply an extension of multiplication:

\begin{equation}
\mathbf{A}^n \equiv \big\{ (a_i)^n \big\} 
\end{equation}

It becomes useful to define a vector space over the field $\mathbb{G}$ which the data objects belong to, using the field $\mathbb{D}$ as elements of the vector space. Accordingly, for addition and multiplication:

\[
\begin{array}{r l}
\circ&: \mathbb{D} \times \mathbb{D} \to \mathbb{D} \\
\times&: \mathbb{G} \times \mathbb{D} \to \mathbb{D}
\end{array}
\]

For example, if $\mathbf{X}$ is a set of 10 floating point numbers (floats), then the corresponding data structure is all possible sets of 10 floats. Operations between $\mathbf{X}$ and another set of 10 floats are now defined, and each member of $\mathbf{X}$ is a float.

Using a more complicated example, one can examine $\mathbf{X}$ as a set of 10 lists, each containing 3 floats and 1 modulo number. In this case, $\mathbb{D}$ is all possible combinations of 10 lists each with 3 floats and 1 modulo number, and each member of $\mathbf{X}$ is a list of 3 floats and 1 modulo.

\section{Primed moments}
Suppose that we define the `updated' dataset containing new values as $\mathbf{X}' = \{x_i'\}$ such that:

\begin{equation}
\mathbf{X} \subset \mathbf{X}'
\end{equation}

$\mathbf{X}'$ is now said to have a cardinality of $N'$. The difference in cardinality is worth noting, and is herein referred to as $\Delta$:

\begin{equation}
\Delta \equiv N' - N
\end{equation}

We will also introduce the following shorthand:

\begin{equation}
\sum\limits_{i=N+1}^{N'} \equiv \sum\limits^{\Delta}
\end{equation}

\subsection{Integer $n$}
For integer $n$, the primed moment can be represented in the following form (see Appendix for proof):

\begin{equation}
M_n' = \dfrac{Z}{Z'} \left[(\bar{x} - \bar{x}')^n  + \sum\limits_{k=0}^{n-2} \binom{n}{k} M_{n-k} (\bar{x} - \bar{x}')^k \right] + \dfrac{1}{Z'} \sum\limits^{\Delta}\left[f(x_i) (x_i - \bar{x}')^n\right]
\label{eq:integer}
\end{equation}

where $\binom{n}{k}$ are binomial coefficients. $Z'$ can be determined using only a simple update:

\begin{figure}[t!]
    \centering
    \includegraphics[width=0.4\textwidth]{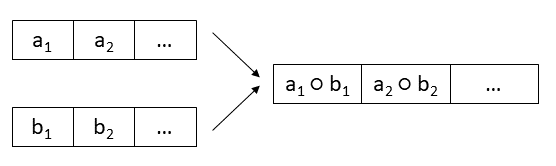}
    \caption{Illustration of the data addition operator $\circ: \mathbb{G} \times \mathbb{G} \to \mathbb{G}$.}
    \label{fig:upper_operations}
\end{figure}

\begin{equation}
Z' = Z + \sum\limits^{\Delta} f(x_i)
\end{equation}

which is computed in $\mathcal{O}(\Delta)$ time. Similarly, $\bar{x}'$ can be computed in $\mathcal{O}(\Delta)$ time:

\begin{equation}
\bar{x}' = \dfrac{Z}{Z'} \bar{x} + \dfrac{1}{Z'} \sum\limits^{\Delta} f(x_i) x_i
\end{equation}

The only points from $\mathbf{X}'$ that are required are the points not in $\mathbf{X}$; viz. the only data required to update the moment is the \emph{new} data. If all moments use the same amount of memory, then the data storage required is only:

\begin{equation}
S_{\Delta} = \textrm{size}(M_n) \cdot (n-1) + \sum\limits^{\Delta} [\textrm{size}(f(x_i)) + \textrm{size}(x_i)]
\end{equation}

This reduced data requirement leads to a computational speed up as well. The computational time taken per update is given by:

\begin{equation}
\dfrac{T}{u} = \mathcal{O}[(n-1)\cdot \Delta]
\end{equation}

One quickly show that by comparing to \eqref{eq:orig_speed} for speed up, one requires at worst:

\begin{equation}
n-1 \lesssim \dfrac{N'}{\Delta}
\end{equation}

This yields a threshold moment value:

\begin{equation}
n_{\textrm{thr.}} \sim \dfrac{N'}{\Delta} + 1
\end{equation}

For this value of $n$ or greater, \eqref{eq:integer} produces the primed moment slower than one could achieve using the original dataset; the speed-up is greatest for small updates to the dataset. However, in all cases, this method uses significantly less data storage.

\subsection{Non-integer $n$}
For non-integer values of $n$, one can use the fractional continuation of the binomial expansion:

\begin{equation}
\begin{array}{r l}
M_n' &= \dfrac{Z}{Z'} \bigg[(\bar{x} - \bar{x}')^n  + \displaystyle\sum\limits_{k=0}^{n-2} \binom{n}{k} M_{n-k} (\bar{x} - \bar{x}')^k \\
&\hspace{50pt} + \displaystyle\sum\limits_{k=n+1}^{\infty} \binom{n}{k} M_{n-k} (\bar{x} - \bar{x}')^k \bigg] + \dfrac{1}{Z'} \displaystyle\sum\limits^{\Delta}\left[f(x_i) (x_i - \bar{x}')^n\right]
\end{array}
\label{eq:noninteger}
\end{equation}

Although it appears that the infinite sum is singular for $\bar{x}' = \bar{x}$, we have already assumed \emph{a priori} that $\bar{x}' \neq \bar{x}$. As $\mathbf{X}$ exists in a vector space, it has a well defined norm. We expect that the infinite sum is Abel convergent for $|x_i - \bar{x}| > 1$, where $|A|$ denotes the Euclidean norm of the dataset $A$ on the vector space established in Section \ref{sec:operations}:

\[
|A| = \left\{\sqrt{a_i \times a_i}\right\}
\]

This condition is therefore related to the standard deviation of the original dataset. In short, we posit that for a set of data with a low standard deviation ($\sqrt{M_2}$), the infinite sum in \eqref{eq:noninteger} should converge. Therefore, one can truncate the sum so as to keep enough terms to yield the answer to a desired level of numerical accuracy:

\begin{equation}
\begin{array}{r l}
M_n' &\approx \dfrac{Z}{Z'} \bigg[(\bar{x} - \bar{x}')^n  + \displaystyle\sum\limits_{k=0}^{n-2} \binom{n}{k} M_{n-k} (\bar{x} - \bar{x}')^k \\
&\hspace{50pt} + \displaystyle\sum\limits_{k=n+1}^{n^*} \binom{n}{k} M_{n-k} (\bar{x} - \bar{x}')^k \bigg] + \dfrac{1}{Z'} \displaystyle\sum\limits^{\Delta}\left[f(x_i) (x_i - \bar{x}')^n\right]
\end{array}
\end{equation}

\begin{figure}[t!]
    \centering
    \includegraphics[width=0.4\textwidth]{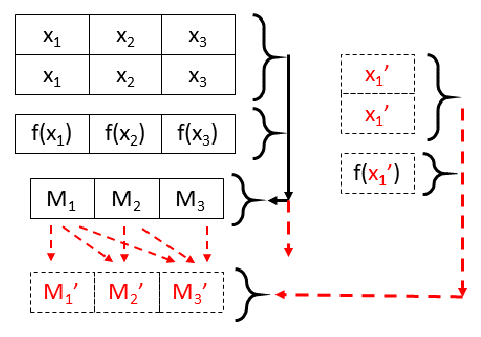}
    \caption{Illustration of the new type of algorithm one can employ to calculate moments and metrics. Only the data to be appended and the previous moments prior to data appending are required in storage.}
    \label{fig:new_alg}
\end{figure}

where $n^*$ is the cutoff moment. Therefore, to numerical accuracy, one requires for data storage:

\begin{equation}
S_{\Delta} = \textrm{size}(M_n) \cdot (n^*-1) + \sum\limits^{\Delta}[\textrm{size}(f(x_i)) + \textrm{size}(x_i)]
\end{equation}

and the computational time taken per update is given by:

\begin{equation}
\dfrac{T}{u} = \mathcal{O}[(n^*-1)\cdot \Delta]
\end{equation}

\section{Weighted metrics}
Suppose that we define the metric $W$ as:

\begin{equation}
W = \dfrac{1}{Z} \sum\limits_{i=0}^{N} f(x_i) g(x_i)
\end{equation}

where $g(x_i)$ is an arbitrary function. Then, one can Taylor expand the function $g(x_i)$ of the dataset around the mean value:

\begin{equation}
W = \sum\limits_{n=0}^{\infty} c_n M_n
\end{equation}

where $c_n$ are Taylor coefficients.

\subsection{Updating metrics}
If one updates the function $g \to g'$, and the dataset $\mathbf{X} \to \mathbf{X}'$ then:

\begin{equation}
W' = \dfrac{Z}{Z'} \sum\limits_{n=0}^{\infty} \sum\limits_{k=0}^{n} \left[ c_n' \binom{n}{k} M_{n-k} (\bar{x} - \bar{x}')^k \right] + \dfrac{1}{Z'} \sum\limits^{\Delta} f(x_i) g'(x_i)
\label{eq:updating}
\end{equation}

where $c_n'$ denotes the new Taylor coefficients. This leads to a very powerful tool; provided that the moments of the original dataset are known it is possible for us to generate any arbitrary metric.

It is worth noting that if one rearranges for the change between $W'$ and $W$:

\begin{equation}
(W' - W) = \dfrac{Z}{Z'} \sum\limits_{n=0}^{\infty} \sum\limits_{k=1}^{n} \left[ c_n' \binom{n}{k} M_{n-k} (\bar{x} - \bar{x}')^k \right] + \dfrac{1}{Z'} \sum\limits^{\Delta} f(x_i) g'(x_i)
\end{equation}

In principle, if one had $g'(x_i) \equiv g(x_i)$, then this method is not particularly useful; we are essentially Taylor expanding at two different points (because the mean shifts), making our life much harder. This is not immediately intuitive; our expansion for $g$ employed here is always around the mean point of the dataset, rather than a free choice.

However, if $W$ is unknown and $\{M_n\}$ are known, then \eqref{eq:updating} is useful.

One can swap the order of summation on the first term. If one does so:

\begin{figure}[t!]
    \centering
    \begin{subfigure}[t]{0.45\textwidth}
    \centering
    \includegraphics[width=\textwidth]{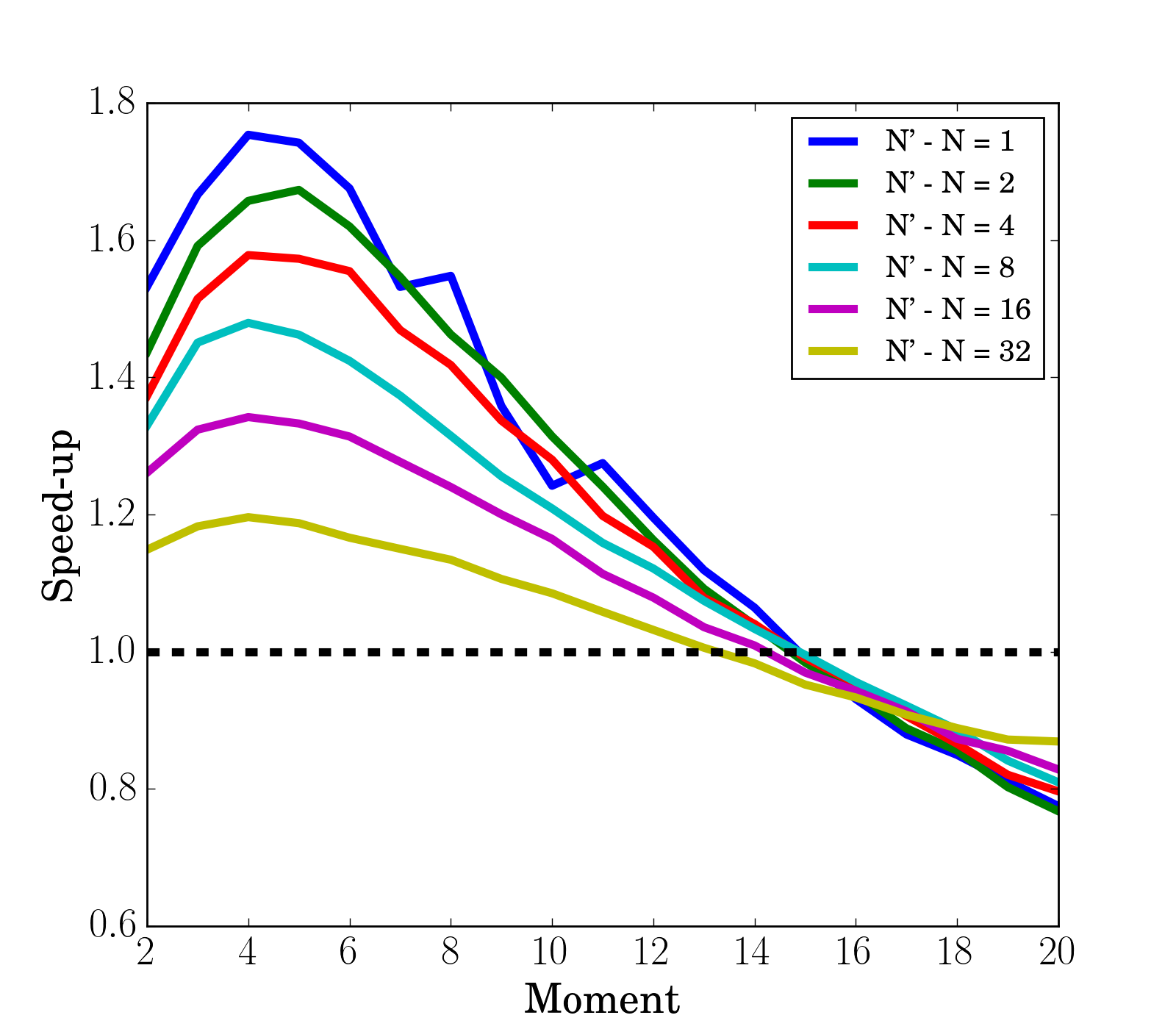}
    \caption{16 float set; speed-up decreases with increasing order of moment, leading to slower performance after roughly the $14^{\textrm{th}}$ moment.}
    \end{subfigure} \hspace{10pt}
    \begin{subfigure}[t]{0.45\textwidth}
    \centering
    \includegraphics[width=\textwidth]{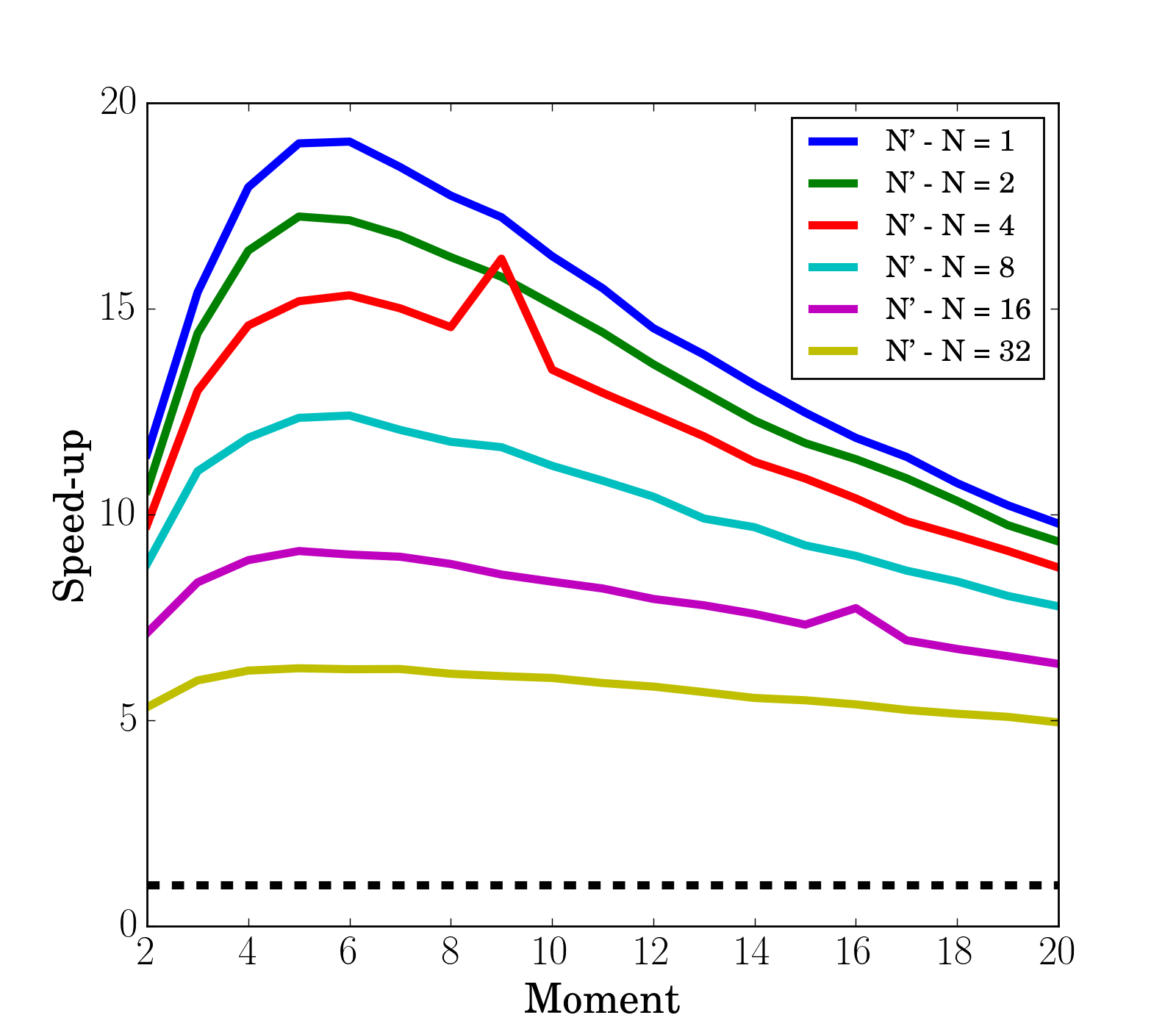}
    \caption{256 float set. Speed-up decreases with increasing order of moment, however the larger set size leads to large speed up everywhere; speed peaks at roughly 19 times faster for the $5^{\text{th}}$ moment.}
    \end{subfigure}
    \caption{Two different datasets (16 floats and 256 floats) from which the $2^{\text{nd}}$ to $20^{\text{th}}$ moments were calculated. Each solid line represents a different value of $N' - N$ (the number of floats added in the update).} 
    \label{fig:floats}
\end{figure}

\[
\renewcommand{\arraystretch}{3}
\begin{array}{r l}
\text{sum} &= \dfrac{Z}{Z'} \displaystyle\sum\limits_{k=0}^{\infty} \left[ (\bar{x} - \bar{x}')^k \sum\limits_{n=k}^{\infty} c_n' \binom{n}{k} M_{n-k}\right] \\
&= \dfrac{Z}{Z'} \displaystyle\sum\limits_{k=0}^{\infty} \left[(\bar{x} - \bar{x}')^k \sum\limits_{n=0}^{\infty} c_{k+n}' \binom{k+n}{k} M_{n} \right]
\end{array}
\]

But as the binomial coefficients are monotonically increasing (at roughly $k^n$), the inside sum is not necessarily convergent. To guarantee Abel convergence, as $M_n$ should be monotonically increasing also, $c_{k+n}$ must satisfy:

\begin{equation}
\left|\sum\limits_{n=0}^{\infty} c_{k+n}\right| < \infty
\end{equation}

This requirement means that we can only update $W$ via this method if the function $g(x)$ has a convergent Taylor series. In such a case, one can use a cutoff for the first sum over $n$ in \eqref{eq:updating}. Then:

\begin{equation}
W' \approx \dfrac{Z}{Z'} \sum\limits_{n=0}^{n^*} \sum\limits_{k=0}^{n} \left[ c_n' \binom{n}{k} M_{n-k} (\bar{x} - \bar{x}')^k \right] + \dfrac{1}{Z'} \sum\limits^{\Delta} f(x_i) g(x_i)
\end{equation}

The first term is iteratively calculated in $\mathcal{O}(n^*!)$ time (such that one reaches numerical accuracy), while the second term is calculated in $\mathcal{O}(\Delta)$ time.

\begin{figure}[t!]
    \centering
    \begin{subfigure}[t]{0.45\textwidth}
    \centering
    \includegraphics[width=\textwidth]{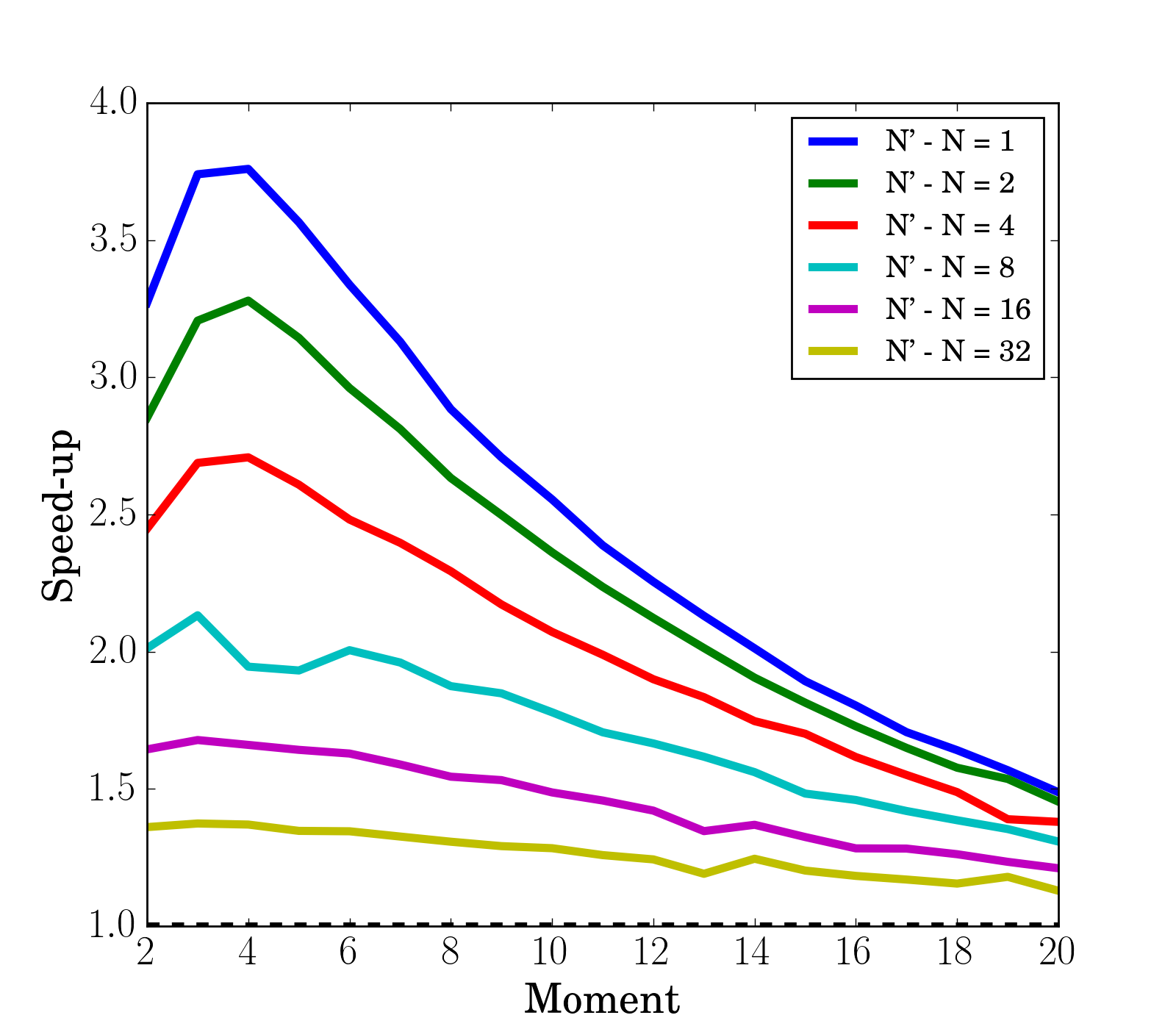}
    \caption{4 sets of 16 floats; speed-up decreases with increasing order of moment, but is greater than 1 for $n \leq 20$.}
    \end{subfigure} \hspace{10pt}
    \begin{subfigure}[t]{0.45\textwidth}
    \centering
    \includegraphics[width=\textwidth]{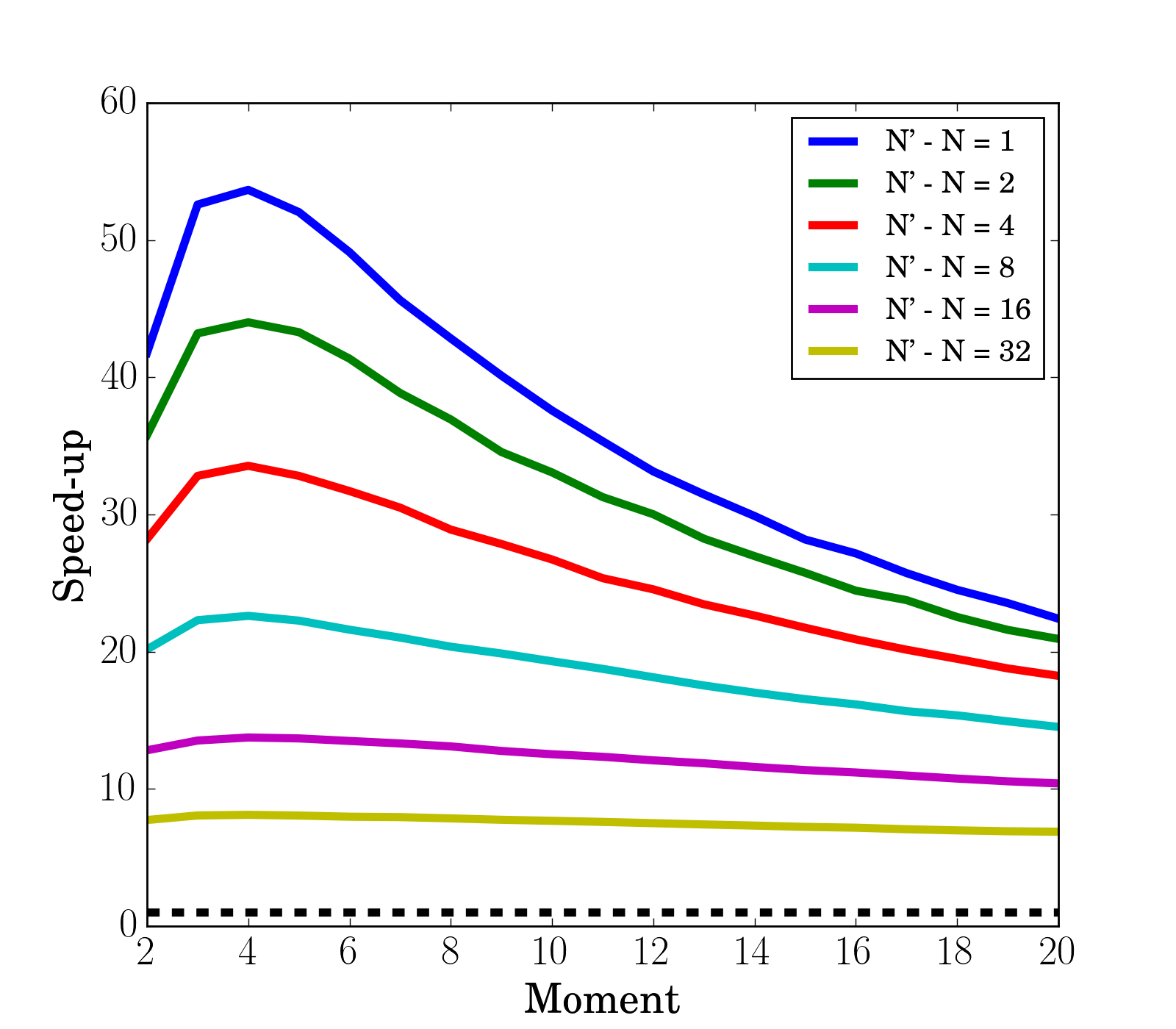}
    \caption{4 sets of 256 floats. Speed-up decreases with increasing order of moment. However, the larger set size leads to large speed up everywhere; speed peaks at roughly 54 times faster for the $2^{\text{nd}}$ moment.}
    \end{subfigure}
    \caption{Two different datasets (4 sets of 16 floats and 4 sets of 256 floats) from which the $2^{\text{nd}}$ to $20^{\text{th}}$ moments were calculated. Each solid line represents a different value of $N' - N$ (the number of floats added in the update).} 
    \label{fig:list}
\end{figure}

\section{Demonstration}
Here, moments of datasets are compared using Python. Python forms a nice framework for adding data structures as one can freely redefine the addition and multiplication operators on objects of an arbitrary class. This makes the formalism from Section \ref{sec:operations} easy to code, allowing us to simply add and multiple entire datasets as if they were just scalars.

\subsection{Single precision floating point}
\label{sec:float}
For single precision floating points with boundaries at 0 and 1:

\begin{equation}
\mathbf{X} = \big\{z : z \in [0,1], z \in \mathbb{R}\big\}
\end{equation}

Two sets of floats, a 16 float set and a 256 float set were used to compute the original integer moments. Then, extra points were added to each set, and the new integer moments were calculated. For each speed test, a new, random set of floats were initialised using the \verb|numpy.random| subpackage.

The scalar function $f(x_i)$ is set to be a random map, such that each piece of data $x_i$ points to a random float. The random floats are reset for each speed test, just as the dataset $\mathbf{X}$ is.

The speed-up was calculated by measuring the time taken to calculate the moment using the full dataset $\mathbf{X}'$, and by using the update method given by \eqref{eq:integer}. For each value of $N' - N$ and moment, the \verb|timeit| package was used to measure 100 runs of the algorithm.

As expected, the speed-up is less than 1 for high moments when the number of data points added ($N' - N$) is large with respect to the size of the original dataset ($N$). One finds that the maximum speed up (for the lowest order moment, $n = 2$) is close to the theoretical maximum of a factor 16 speed up between the $N = 16$ and the $N = 128$ set, as shown in Figure \ref{fig:floats}.

\subsection{List of single floats \label{sec:list}}
We now examine a list of single floats. Using boundaries at 0 and 1 for each float, and setting the list length to 4:

\begin{equation}
\begin{array}{r l}
\mathbf{X} &= \big\{x_i : i \in [0,1,2,3] \big\} \\
x_i &= \big\{z : z \in [0,1], z \in \mathbb{R} \big\}
\end{array}
\end{equation}

The minimum update (in terms of $N \to N + 1$) now adds 4 floats to the dataset. In principle, this could be reformulated using the same analysis as in Section \ref{sec:float} but updating the dataset in groups of 4. As such, we expect that the speed-up will look very similar, but will be affected by the 4 fold global increase in datasize.

Again, we use \verb|numpy.random| to generate a random set of floats, and $f(x_i)$ is a random scalar map. As shown in Figure \ref{fig:list}, the speed-up for 4 sets of 16 floats is fairly similar to that observed with a single set of 16 floats. However, when we increase to 4 sets of 256 floats, the speed up is much greater than the case where we examine a single set of 256 floats (see Figure \ref{fig:floats}). Again, one finds that the maximum speed up (for the lowest order moment, $n = 2$) is close to the theoretical maximum of a factor 16.

\section{Conclusion}
In conclusion, this algorithm and method allows one to swiftly update moments and arbitrary metrics of datasets using less memory, and less data storage. The resultant code is lightweight, and can easily be implemented using class-oriented programming in a language of the readers' choice (only Python results are shown here).

This algorithm is released as copyleft under the GNU GPL v3 license.

{\small
\bibliographystyle{unsrt}
\bibliography{biblio}}

\begin{appendices}
\renewcommand\thesection{:}
\renewcommand{\theequation}{A-\arabic{equation}}

\section{Integer $n$ updates (proof)}
\label{app:integer}
\textbf{To be demonstrated:}

\[
M_n' = \displaystyle \dfrac{Z}{Z'} \left[(\bar{x} - \bar{x}')^n + \sum\limits_{k=0}^{n-2} \binom{n}{k} M_{n-k} (\bar{x} - \bar{x}')^k \right] + \dfrac{1}{Z'} \left[\sum\limits^{\Delta} f(x_i) (x_i - \bar{x}')^n\right] 
\]

where $Z'$ is given by:

\[
Z' = Z + \sum\limits^{\Delta} f(x_i)
\]

and $\bar{x}'$ is given by:

\[
\bar{x}' = \dfrac{Z}{Z'} \bar{x} + \dfrac{1}{Z'} \sum\limits^{\Delta} f(x_i) x_i 
\]

\begin{proof}
From \eqref{eq:moment}, one finds that after update:

\begin{equation}
M_n' = \dfrac{1}{Z'} \sum\limits_{i=1}^{N'} f(x_i) (x_i - \bar{x}')^n
\label{eq:mnupdate}
\end{equation}

where $\bar{x}'$ is given by:

\begin{equation}
\bar{x}' = \dfrac{1}{Z'} \sum\limits_{i=1}^{N'} f(x_i) x_i
\label{eq:xbar}
\end{equation}

First, if one examines $Z'$:

\[
\begin{array}{r l}
Z' &= \sum\limits_{i=1}^{N'} f(x_i) \\
& = \sum\limits_{i=1}^{N} f(x_i) + \sum\limits^{\Delta} f(x_i) \\
& = Z + \sum\limits^{\Delta} f(x_i)
\end{array}
\]

such that the first term is the original value of the normalising function, and the second term represents the change in the value from adding new data.

If we split \eqref{eq:xbar} into a sum over $1$ to $N$, and $N+1$ to $N'$:

\[
\begin{array}{r l}
\bar{x}' &= \dfrac{Z}{Z'} \left[\sum\limits_{i=1}^{N} \dfrac{1}{Z} f(x_i) x_i\right] + \dfrac{1}{Z'} \sum\limits^{\Delta} f(x_i) x_i \\
&= \dfrac{Z}{Z'} \bar{x} + \dfrac{1}{Z'} \sum\limits^{\Delta} f(x_i) x_i 
\end{array}
\]

One finds that the following holds true:

\[
\begin{array}{r l}
(x_i - \bar{x}')^n &= ([x_i - \bar{x}] + [\bar{x} - \bar{x}'])^n \\
&= \sum\limits_{k=0}^{n} \binom{n}{k} [x_i - \bar{x}]^{n-k} (\bar{x} - \bar{x}')^k
\end{array}
\]

where we have performed a binomial expansion of modified form of $(x_i - \bar{x}')^n$. By substituting the above into \eqref{eq:mnupdate}:

\[
\renewcommand{\arraystretch}{3}
\begin{array}{r l}
M_n' &= \displaystyle\dfrac{1}{Z'} \left[\sum\limits_{i=1}^{N} f(x_i) (x_i - \bar{x}')^n\right] + \dfrac{1}{Z'} \left[\sum\limits^{\Delta} f(x_i) (x_i - \bar{x}')^n\right] \\
&= \displaystyle \sum\limits_{i=1}^{N} \sum\limits_{k=0}^{n} \binom{n}{k} [x_i - \bar{x}]^{n-k} (\bar{x} - \bar{x}')^k + \dfrac{1}{Z'} \left[\sum\limits^{\Delta} f(x_i) (x_i - \bar{x}')^n\right] \\
&= \displaystyle \dfrac{Z}{Z'} \sum\limits_{k=0}^{n} \left[\binom{n}{k} M_{n-k} (\bar{x} - \bar{x}')^k \right] + \dfrac{1}{Z'} \left[\sum\limits^{\Delta} f(x_i) (x_i - \bar{x}')^n\right] \\
&= \displaystyle \dfrac{Z}{Z'} \left[\binom{n}{n} M_0 (\bar{x} - \bar{x}')^n + \sum\limits_{k=0}^{n-2} \binom{n}{k} M_{n-k} (\bar{x} - \bar{x}')^k \right] + \dfrac{1}{Z'} \left[\sum\limits^{\Delta} f(x_i) (x_i - \bar{x}')^n\right] \\
&= \displaystyle \dfrac{Z}{Z'} \left[(\bar{x} - \bar{x}')^n + \sum\limits_{k=0}^{n-2} \binom{n}{k} M_{n-k} (\bar{x} - \bar{x}')^k \right] + \dfrac{1}{Z'} \left[\sum\limits^{\Delta} f(x_i) (x_i - \bar{x}')^n\right] \qedhere
\end{array}
\]

where we noted that $M_1 = 0$. 
\end{proof}
\end{appendices}

\end{document}